\documentclass[a4paper,11pt]{article}

\usepackage{contribution}
\newcommand{\weblink}[2][]{%
    \ifthenelse{\equal{#1}{}}%
    {\textnormal{\url{#2}}}%
    {\textnormal{\href{#2}{#1}}}%
}

\def\beq{\begin{equation}}
\def\eeq#1{\label{#1}\end{equation}}
\def\eeqn{\end{equation}}

\def\beqa{\begin{eqnarray}}
\def\eeqa#1{\label{#1}\end{eqnarray}}
\def\eeqan{\end{eqnarray}}

\let\bar=\overbar

\def\Dslash{\not{\hbox{\kern-4pt $D$}}}
\def\dslash{\not{\hbox{\kern-2pt $\del$}}}

\def\msb{{\bar{\ssstyle M \kern -1pt S}}}

\newcommand{\contribution}[7][]{%
  \clearpage
  \thispagestyle{plain}
  \ifthenelse{\equal{#1}{}}
  {\hypersetup{pdftitle={#2}}}
  {\hypersetup{pdftitle={#1}}}
  \hypersetup{pdfauthor={{#3} {#4}}}
  {\centering\normalfont\LARGE\bfseries\sffamily #2 \par\nobreak}
  \lhead{}
  \chead{%
    \textit{\footnotesize XIV International Conference on Hadron Spectroscopy
      (\weblink[\textit{hadron2011}]{http://www.hadron2011.de}), 13-17 June 2011, Munich, Germany}%
  }
  \rhead{}
  \bigskip
  \begin{center}
    {#3} {#4}\ifthenelse{\equal{#6}{}}{}{\footnote{\weblink[#6]{mailto:#6}}}
    \ifthenelse{\equal{#7}{}}{}{#7} \\
    \textit{#5}
  \end{center}
  \bigskip
}

\renewcommand{\abstract}[1]{%
  \begin{center}
    \begin{minipage}{0.85\textwidth}
      \begin{footnotesize}
        #1
      \end{footnotesize}
    \end{minipage}
  \end{center}
  \bigskip
}

\begin{document}

% % % % % % % % % % % % % % % % % % % % % % % % % % % % % % % % % % % % % % % % %
% your proceedings
%%%%%%%%%%%%%%%%%%%%%%%%%%%%%%%%%%%%%%%%%%%%%%%%%%%%%%%%%%%%%%%%%%%%%%%%%%%%%%%%%
%
% template for hadron2011 contribution
%
% please do not rename this file
%
% to create document run
%
%     pdflatex hadron2011.tex
%
%%%%%%%%%%%%%%%%%%%%%%%%%%%%%%%%%%%%%%%%%%%%%%%%%%%%%%%%%%%%%%%%%%%%%%%%%%%%%%%%%
{  % do not remove

%%%%%%%%%%%%%%%%%%%%%%%%%%%%%%%%%%%%%%%%%%%%%%%%%%%%%%%%%%%%%%%%%%%%%%%%%%%%%%%%%
% please define your macros here

%
%%%%%%%%%%%%%%%%%%%%%%%%%%%%%%%%%%%%%%%%%%%%%%%%%%%%%%%%%%%%%%%%%%%%%%%%%%%%%%%%%

%%%%%%%%%%%%%%%%%%%%%%%%%%%%%%%%%%%%%%%%%%%%%%%%%%%%%%%%%%%%%%%%%%%%%%%%%%%%%%%%%
% define title, author, and address
% contribution[short title]{title}{author first name}{author last name}{author address}{author email}{collaboration}
% the short title will appear in the page headers and the TOC of the book of proceedings
% the last two arguments may be left empty
\contribution[]  % short title (optional)
{Structure of scalar mesons \\ and the Higgs sector of strong interaction}  % title
{Martin}{Schumacher}  % first and last name of author
{Zweites Physikalisches Institut\\
  der Universit\"at G\"ottingen \\
  D-37077 G\"ottingen, Germany}  % author address
{mschuma3@gwdg.de}  % author email optional
{}  % collaboration (optional)
%
%%%%%%%%%%%%%%%%%%%%%%%%%%%%%%%%%%%%%%%%%%%%%%%%%%%%%%%%%%%%%%%%%%%%%%%%%%%%%%%%%

%%%%%%%%%%%%%%%%%%%%%%%%%%%%%%%%%%%%%%%%%%%%%%%%%%%%%%%%%%%%%%%%%%%%%%%%%%%%%%%%%
% abstract
\abstract{
The $\sigma$ meson may be considered as the Higgs boson of strong interaction.
While the observation of the electroweak Higgs boson is the  primary goal
in ongoing experiments at the LHC, the $\sigma$ meson is by now well studied
both  as an on-shell particle and as a virtual particle while being part of 
the constituent quark. This makes it timely to give an overview of the present
status of the Higgs sector of strong interaction which includes the scalar
mesons $\sigma(600)$, $\kappa(800)$, $f_0(980)$ and $a_0(980)$ together with
the pseudo Goldstone bosons $\pi$, $K$ and $\eta$.
}%
%
%%%%%%%%%%%%%%%%%%%%%%%%%%%%%%%%%%%%%%%%%%%%%%%%%%%%%%%%%%%%%%%%%%%%%%%%%%%%%%%%%

%%%%%%%%%%%%%%%%%%%%%%%%%%%%%%%%%%%%%%%%%%%%%%%%%%%%%%%%%%%%%%%%%%%%%%%%%%%%%%%%%
% main text
% for short contributions sections are optional
\section{Introduction}

Scalar mesons below 1 GeV together with the pseudo Goldstone bosons $\pi$, $K$
and $\eta$ may be considered as the Higgs sector of strong interaction. 
While the EW Higgs boson up to now 
appears to escape experimental observation in the
ongoing LHC experiments \cite{schumacher11} the strong counterpart, 
the $\sigma$ meson is by now
well studied both as on-shell particle and as a virtual particle while being
part of the constituent quark. The latter observation has been facilitated
through Compton scattering by the proton in an experiment carried out at
MAMI (Mainz) published in 2001 \cite{galler01,wolf01}. 
In this experiment it has been shown that
the scalar $t$-channel makes a strong contribution to the Compton scattering 
amplitude, being  successfully  represented in terms of a 
$t$-channel pole located at $m^2_\sigma$ where  $m_\sigma$ is the bare 
mass of the $\sigma$ meson, determined in this experiment to be $\sim 600$ MeV.
Inspite of this great success the physical interpretation of the experiment 
remained uncertain because an explicit $\sigma$ meson is a strongly unwanted
particle in chiral perturbation theory. This led to an unnecessary
delay, because  
a detailed theoretical investigation was required  extending until   2010,
when it was shown that the $t$-channel
pole at $m^2_\sigma$ is a well founded concept 
and that the related
$t$-channel amplitude may be understood as being due to
Compton scattering by the $\sigma$ meson while being part of the constituent
quark \cite{schumacher10}. The findings in \cite{schumacher10} were 
extended to include the whole
scalar nonet below 1 GeV in \cite{schumacher11a}. The present work is in part
based on this latter publication where more details may be found.

\section{The doorway model and the structure of scalar mesons}

The scalar nonet below 1 GeV cannot be 
understood in terms of  flavor structures as provided by 
$SU(3)_f$ \cite{jaffe76} because of the ordering of the meson masses.
This problem was  solved by introducing tetraquarks $(q\bar{q})^2$
\cite{jaffe76}.  The tetraquark model implies the possibility of
a dissociation
of the kind $(q\bar{q})^2\rightleftarrows (q\bar{q}+q\bar{q})$, leading to 
$q\bar{q}$  as a small structure component. In \cite{schumacher11a} this
small $q\bar{q}$ structure component was interpreted in terms of a doorway
state which serves as the
entrance channel  in a two-photon fusion reaction and is in agreement with the
experimental two-photon widths of the mesons:
\begin{eqnarray}
&& \sigma(600):\quad\, \gamma\gamma\to \frac{u\bar{u}+d\bar{d}}{\sqrt{2}}\to
u\bar{u}d\bar{d}\to\pi\pi, \label{eq1}\\
&&f_0(980):\quad
\gamma\gamma\to\frac{1}{\sqrt{2}}\left(\frac{u\bar{u}+d\bar{d}}
{\sqrt{2}}-s\bar{s}  \right)\to \frac{s\bar{s}(u\bar{u}+d\bar{d})}{\sqrt{2}}
\to \pi\pi,K\bar{K}, \label{eq2}\\
&&a_0(980):\quad
\gamma\gamma\to\frac{1}{\sqrt{2}}\left(\frac{-u\bar{u}+d\bar{d}}
{\sqrt{2}}+s\bar{s}  \right)\to \frac{s\bar{s}(u\bar{u}-d\bar{d})}{\sqrt{2}}
\to \eta\pi,K\bar{K}. \label{eq3}
\end{eqnarray}
The $q\bar{q}$
configuration  of the $a_0(980)$ meson violates isospin conservation. This is
of no problem because we consider the $q\bar{q}$ configuration  only as a small
structure component.

In $t$-channel nucleon Compton scattering the reaction chain
\begin{equation}
\gamma\gamma\to \{\sigma(600), f_0(980), a_0(980)\}\to N\bar{N}
\label{eq4}
\end{equation} 
is considered where the excitation of the $N\bar{N}$ pair is virtual. This
leads to the consequence that the masses of the scalar mesons entering into
(\ref{eq4}) are the bare masses, i.e. the masses for the case of zero particle
decay width. The validity of this concept has been shown in
\cite{schumacher09} where quantitative predictions of electromagnetic
polarizabilities of the nucleon led to excellent agreement 
with experimental data.

\section{Mass prediction for scalar mesons in terms of spontaneous, dynamical
  and explicit symmetry breaking}

In case of pseudoscalar and scalar mesons the following phenomena contribute
to the generation of the masses  of the mesons:\\
(i) The $U(1)_A$ anomaly,\\
(ii) spontaneous or dynamical symmetry breaking,\\
(iii) explicit symmetry breaking leading to non-zero current-quark masses.\\
The $U(1)_A$ anomaly is a gluonic (instanton \cite{thooft86}) 
effect which works on $SU(3)_f$ flavor states
which are completely symmetric in the chiral limit. For pseudoscalar and
scalar mesons this is only the case for the $\eta_0$ flavor state and has the
consequence that  $\eta_0$ has a mass in the chiral limit whereas all the
other pseudoscalar mesons are massless. These latter pseudoscalar mesons form
the octet of Goldstone bosons as depicted in the left panels of Figures
\ref{mexicanhat} and \ref{levelscheme}. 
\begin{figure}[h]
\begin{center}
\includegraphics[width=0.21\linewidth]{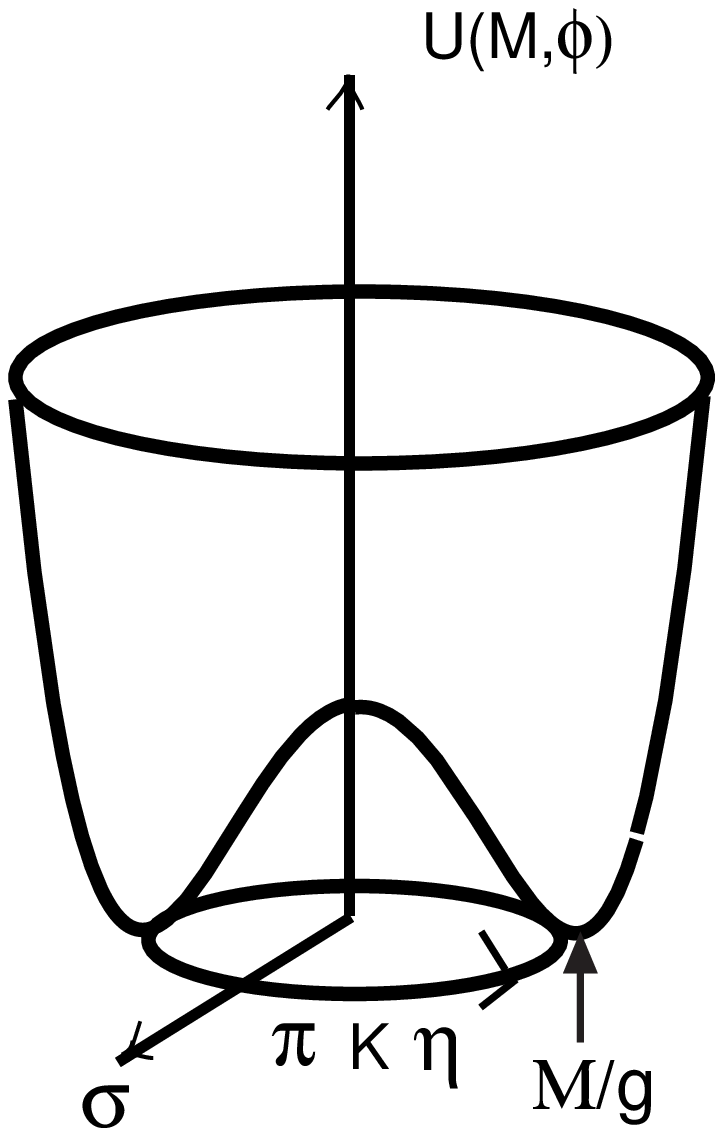}
\includegraphics[width=0.4\linewidth]{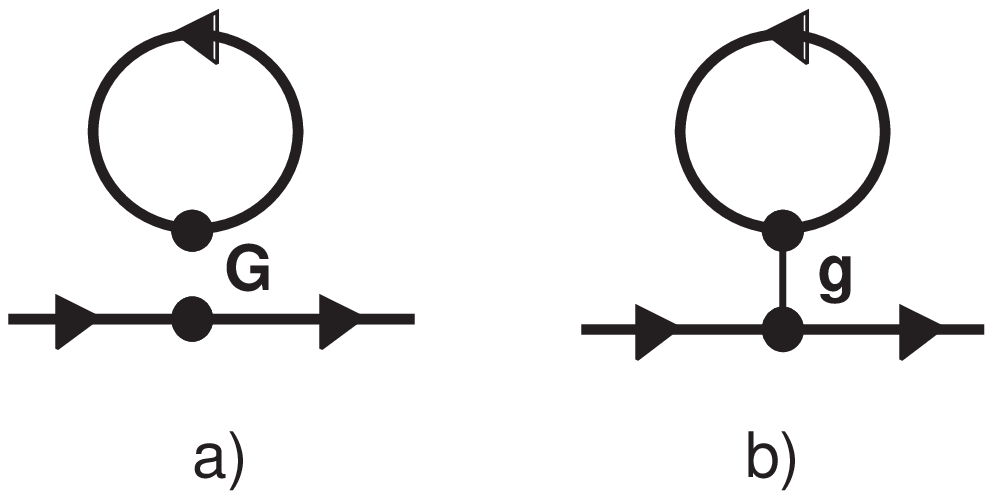}
\end{center}
\caption{\small Left panel: Spontaneous symmetry breaking in the chiral limit (cl)
illustrated by the  L$\sigma$M: In the $SU(2)$ sector there is one 
``strong Higgs boson'', 
the  $\sigma$ meson having a mass of $m^{\rm cl}_\sigma=652$ MeV taking 
part in spontaneous  symmetry breaking, accompanied by an isotriplet
of massless $\pi$ mesons serving as Goldstone bosons. In the $SU(3)$ sector
there are  8 massless Goldstone bosons $\pi$, $K$, $\eta$, and nine 
scalar mesons
$\sigma$, $\kappa$, $f_0$ and $a_0$, all of them 
having the same mass as the $\sigma$ meson in the chiral limit. 
The mass degeneracy is removed
by explicit symmetry breaking.
Right panel: Tadpole graphs of dynamical
symmetry breaking. a) Four fermion version of the Nambu-Jona--Lasinio (NJL)
model, b) bosonized NJL model.}
\label{mexicanhat}
\end{figure}
\begin{figure}[h]
\begin{center}
\includegraphics[width=0.26\linewidth]{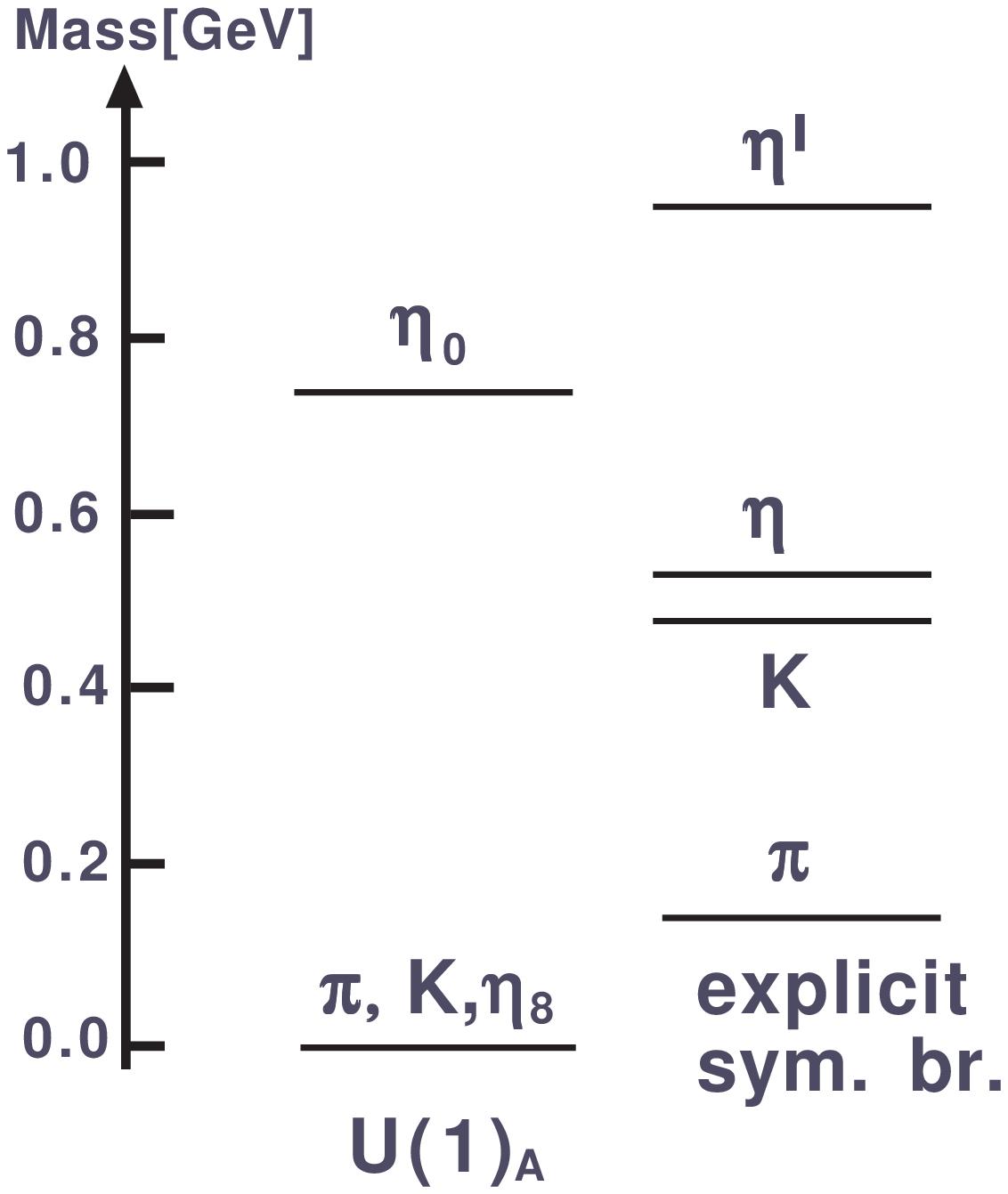}
\includegraphics[width=0.32\linewidth]{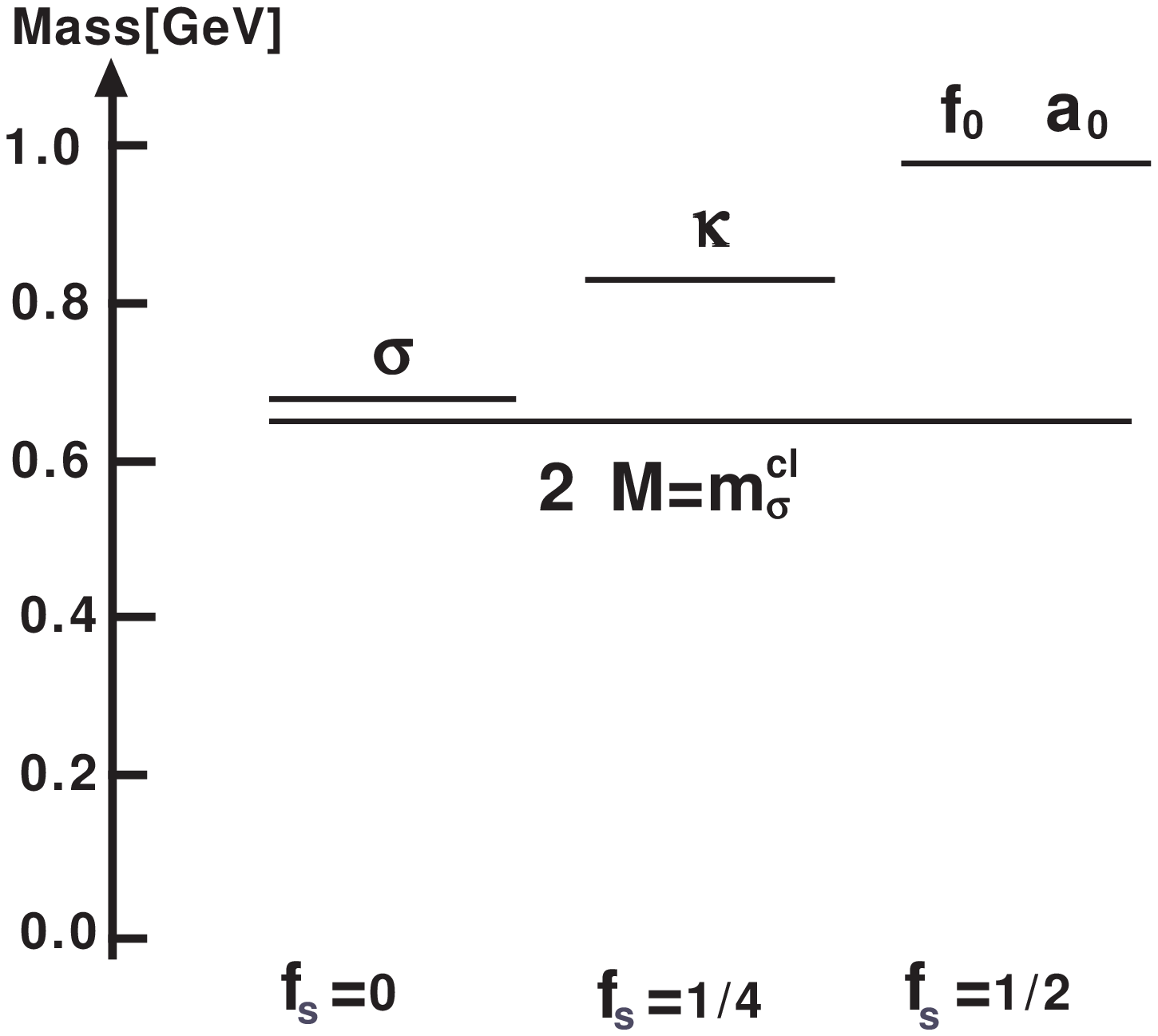}
\end{center}
\caption{\small Left panel: Pseudoscalar mesons after U(1)$_A$ symmetry breaking
  (left column) and after additional explicit symmetry breaking (right
  column). 
Right panel: Masses of the members of the scalar nonet. In  the chiral limit
  all the scalar mesons have the same mass amounting to 
$2M=m^{\rm cl}_\sigma=652$ MeV, where $M$ is the mass of the constituent
  quark in the chiral limit and $m^{\rm cl}_\sigma$ the mass of the $\sigma$
  mesons in the chiral limit (cl). Explicit symmetry breaking shifts the
  masses upward with the fraction $f_s$ of strange quarks in the tetraquark
  structure being the parameter determining the size of the shift.}
\label{levelscheme}
\end{figure}
The left panel of Figure
\ref{mexicanhat} shows the mexican-hat potential where the Goldstone bosons
correspond to the minimum of the potential. The mexican-hat potential
describes spontaneous symmetry breaking in terms of a mass parameter 
$\mu$ and a self-coupling parameter $\lambda$. Since these parameters are
unknown no quantitative prediction of the masses of the constituent quark
and of the scalar mesons is possible. This  is different in the
quark-level linear $\sigma$ model  (QLL$\sigma$M) where the graphs shown in
Figure 1 a) and b) are taken into account. In this way the Delbourgo-Scadron
relation \cite{delbourgo95}
\begin{equation}
M=g\,f_0
\label{del}
\end{equation}
is obtained with $g=2\pi/\sqrt{3}$ being the $\sigma$-quark coupling constant
and $f_0=89.8$ MeV the pion decay constant in the chiral limit. Eq.
(\ref{del}) leads to $m^{\rm cl}=2M=652$ MeV as given in the caption of Figure 
\ref{levelscheme}. Explicit symmetry breaking is described by generalizing the
mass formula valid for the $\sigma$ meson
\begin{equation}
m^2_\sigma=\frac{16\pi^2}{3}f^2_\pi+m^2_\pi
\label{mass1}
\end{equation}
by taking into account the larger fraction of strange quarks in the
$\kappa(800)$ and the $(f_0(980),a_0(980))$ mesons in their tetraquark
structures. This leads to
\begin{eqnarray}
&&m^2_\kappa=\frac{16\pi^2}{3}\frac12(f^2_\pi+f^2_K)+\frac12(m^2_\pi+m^2_K)
\label{mass2}\\
&&m^2_{a_0,f_0} =\frac{16\pi^2}{3}f^2_K+m^2_\eta
\label{mass3}
\end{eqnarray}
where $f_\pi=92.42\pm0.26$ MeV and $f_K=113.0\pm1.0$ MeV. The masses predicted
in this way are $m_\sigma=685$ MeV, $m_\kappa=834$ MeV and $m_{a_0,f_0}= 986$
MeV in close agreement with the experimental data.

\end{document}